\documentclass[11pt,twoside]{article}

\usepackage{edwin}
\usepackage{epsf}
\usepackage{psfig}
\usepackage{lscape}
\usepackage{longtable}
\usepackage{latexsym}
\usepackage{amssymb,amsmath}

\begin{document}
\title{The ADS in the Information Age - Impact on Discovery}   
\author{Edwin A. Henneken\altaffilmark{1}, Michael J. Kurtz, Alberto Accomazzi}
\affil{Harvard-Smithsonian Center for Astrophysics, 60 Garden Street, Cambridge, MA 02138}
\altaffiltext{1}{Corresponding author at: Harvard-Smithsonian Center for Astrophysics, 60 Garden Street, Cambridge, MA 02138, USA. {\it E-mail address}: ehenneken@cfa.harvard.edu (E. Henneken).}
\begin{abstract} 
The SAO/NASA Astrophysics Data System (ADS) grew up with and has been riding the waves of the Information Age, closely monitoring and anticipating the needs of its end-users. By now, all professional astronomers are using the ADS on a daily basis, and a substantial fraction have been using it for their entire professional career. In addition to being an indispensable tool for professional scientists, the ADS also moved into the public domain, as a tool for science education. In this paper we will highlight and discuss some aspects indicative of the impact the ADS has had on research and the access to scholarly publications.

The ADS is funded by NASA Grant NNX09AB39G.
\end{abstract}
\section{Introduction}

Why do scientists publish? First and foremost, because they want to share their findings and further science. Essential to this process is the ability for other people to (efficiently) discover these publications. The process of information discovery has changed dramatically over the last two decades. Remember the days of spending hours in a library, paging through A\&A Abstracts, catalogs and tables of contents? Sometimes a publication had to be retrieved via Inter Library Loan, adding more time to the discovery process. A lengthy discovery process not only means a long journey before finally acquiring enough fuel to set the first step towards your goal, it also means that the discovery process is a significant portion of the cost of employing a scientist. Back then it was virtually impossible to answer the question ``what is the most popular paper on X among people interested in X?'' or to find a set of review papers on this subject (within a reasonable amount of time). Ease of access is therefore essential for efficient information discovery. When the digital revolution of the Information Age culminated in the birth of the Internet, followed by the World Wide Web, the ingredients were there to take information discovery to a new level. At this time, communication started to change from paper to electronic, and this in turn created a fundamental change in society. Information in electronic form resulted in a big shift in ease of access. But just ease of access is not enough. In order to discover information efficiently, you need tools to explore this electronic universe. The combination of ease of access and (powerful) tools for information discovery are central to the process of transferring knowledge. In astronomy, the ADS has been central and pivotal in this digital revolution. 

It is difficult to objectively quantify the absolute impact of the ADS (or any online service, for that matter), but we will highlight a number of facts that will illustrate aspects of the impact of the ADS. Firstly, the ease of access, combined with the powerful query capabilities of the ADS, has had a direct impact on the scientific process in the form of the amount of time gained that researchers otherwise would have spent going to a library, physically finding an article, Xeroxing it and returning to their office. Also, access through the World Wide Web means that communities that historically had little or no access to the scholarly literature, now have access to at least basic meta data, scanned articles, Open Access literature and full text through e-prints in the arXiv repository (with which the ADS is synchronized every night). Thirdly, by diversifying its holdings, the ADS provides the astronomy community not only with the essential core journals, but also with publications from the ever-expanding periphery. Fields that seemed to have no overlap with astronomy and astrophysics in the past, suddenly become relevant and core journals for those fields start to have content that should be available to astronomers and astrophysicists. Next, through its digitization efforts, the ADS has created access to rare and historical publications. Finally, another measure of impact is the fact that the existence of the ADS, and other electronic resources, has had a direct influence on the publication process itself. We will visit these various modes of impact in more detail in the following sections. 
\section{Impact of the ADS on astronomy}
\subsection{Efficiency of Astronomical Research}
It seems reasonable to assume that an increased efficiency in discovering information will translate into researchers being better informed because, among other things, they will get exposed to a broader range of information sources per search effort. To what level researchers are informed will most certainly influence the quality of their research. The increase in efficiency, using the ADS, is most dramatic for more complex, but still realistic queries. For example, finding review papers on a given subject, or the most popular papers among people also interested in a given subject is a matter of just seconds, using the ADS ``Topic Search''. Doing this type of literature research in a traditional, ``paper'' library would easily take hours. But also compared to other electronic resources, using the ADS will result in increased efficiency and better results. The increase in efficiency becomes even more pronounced in those cases where the ADS provides links to additional sources of information (like SIMBAD, NED, VizieR and on-line data).

In order to quantify the increase in efficiency, \citet{kurtz00}~developed a metric, based on the concept of equivalent research time gained. The physical retrieval of an article (``overhead'') was estimated to be 15 minutes on average, in the ``paper age''. Using the ADS virtually removes this overhead, therefore resulting in a gain of roughly 15 minutes research equivalent time when an article is downloaded. \citet{kurtz00}~further estimate that downloading an abstract, citation list or reference list gains one third of the time of an entire article (5 minutes). Based on the worldwide combined ADS logs from March 1999, \citet{kurtz00}~ found that the impact of the ADS on astronomy is 333 FTE (Full Time Equivalent, 2000 hour) research years per year. In a later study \citet{kurtz05a}~found that, based on the 2002 ADS usage logs, an increase in efficiency of astronomical research by 736 FTE researcher equivalent years, or about 7\% of all research done in astronomy. If we apply this to the current logs (extrapolating from the Jan 2011 logs to a full year) and just the four main astronomy journals ({\it ApJ}, {\it A\&A}, {\it MNRAS}, {\it AJ}), we arrive at a number of 985 FTE research years per year. Whatever the exact interpretation of an FTE researcher year is, the impact of the ADS on astronomy, and science in general, is clearly substantial. 

\begin{figure}[!ht]
  \plotone{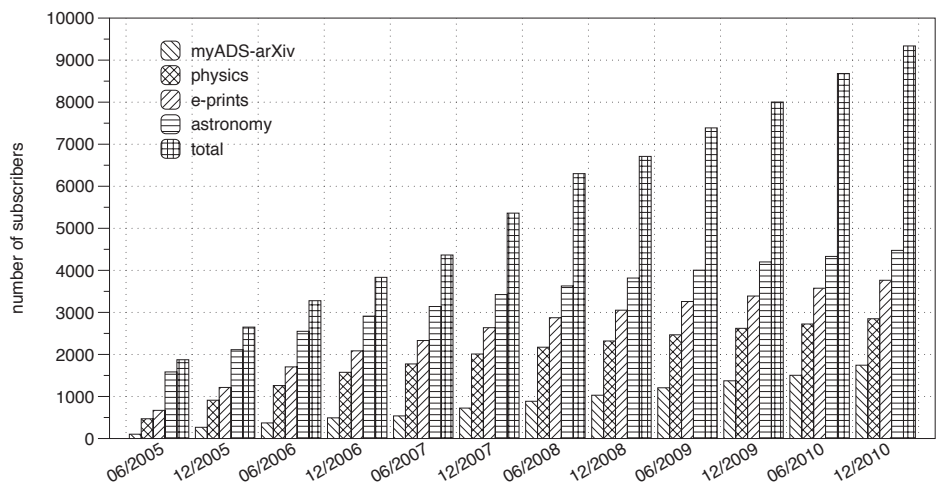}
  \caption{The number of myADS and myADS-arXiv users over time}
  \label{fig:myADSusers}
\end{figure}

Another example of contributions to increased efficiency are the services {\it myADS} and {\it myADS-arXiv}, providing the scientific community with a one stop shop for staying up-to-date. In 2003 the ADS introduced a notification service, {\it myADS}, which uses sophisticated queries (2nd order operators) to give users a powerful tool for staying current with the latest literature in their sub fields of astronomy and physics. The {\it myADS-arXiv} service (``daily myADS'') provides a powerful and unique filter on the enormous amount of bibliographic information added to the ADS on a daily basis, as it gets synchronized with arXiv. In essence {\it myADS-arXiv} is a tailor-made, open access, virtual journal (see \citet{henneken07a}). The {\it myADS} services automate an obvious need for most scientists: answering questions like ``Who is citing my papers?'' and ``What are recent, most popular and most cited papers in my field?''. Automating these queries and providing an alert service saves time. The popularity of these services is reflected in the steady growth of the number of {\it myADS} and {\it myADS-arXiv} users (see figure~\ref{fig:myADSusers}). 

\subsection{Worldwide Access and Sociological Impact}
\begin{figure}[!ht]
  \plotone{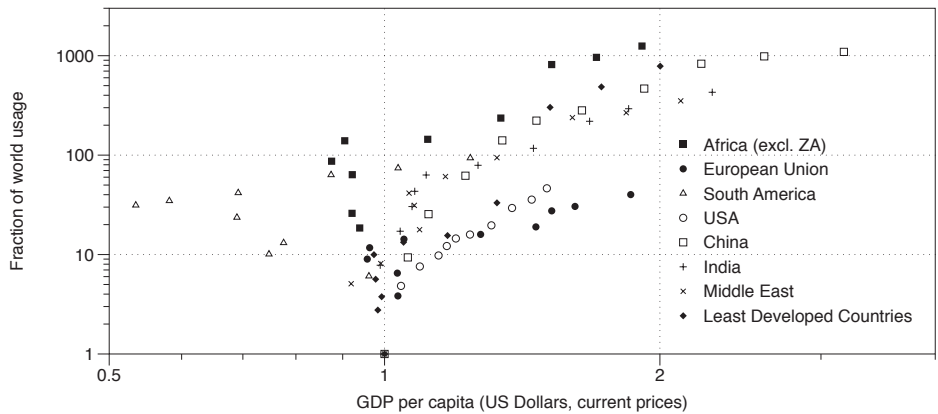}
  \caption{The fraction of world usage for a number of regions as a function of GDP per capita for that region. Both quantities have been normalized by their value in 1997}
  \label{fig:usageGPC}
\end{figure}

In 2002, the Harvard-Smithsonian Center for Astrophysics Visiting Committee reported that the ADS ``{\it empowered astronomy research in underdeveloped countries and small institutions}'' (From Report of the CfA Visiting Committee, 2002). In November of 2005, the United Nations General Assembly commended the ADS for ``the mirror sites of the NASA-funded Astrophysics Data System (ADS). . . had been enthusiastically accepted by the scientific community and had become {\it important assets for developing countries} . . . '' (excerpt from~\citet{UN05}). The ADS has been instrumental in helping to bridge the ``Digital Divide'' (see e.g.~\citet{ITU07}) for astronomical research. In \citet{henneken09}~we showed that increased Internet access, in particular in Least Developed Countries (UN definition), has resulted in increased ADS usage. In that publication we examined readership in a particular region as a function of GDP per capita (GPC), because science and technology depend heavily on available budgets. Figure~\ref{fig:usageGPC} (taken from that paper) shows the relation between normalized GPC and normalized usage for a specific region. The definition for the region ``Least Developed Countries'' was taken from~\citet{UN08}. The numbers have been normalized with respect to the 1997 level, so the diagram shows a relative growth with respect to 1997. The general evolution in this diagram is up and to the right, as time progresses. The data used to construct figure~\ref{fig:usageGPC} were taken from the ADS logs,~the ``Earth Trends Database'' (\citet{WRI08}), the ``World Economic Outlook'' (\citet{IMF08}) and the ``World Statistics Pocketbook - Least Developed Countries'' (\citet{UN07}).

Our logs show that growth in world usage is clearly driven by regions with the biggest potential. High- and middle-income regions have reached a saturation level in the density of Internet users, causing normalized ADS usage to increase at a slower rate. Clearly, the biggest potential is in low-income regions. There is a rapid increase in Internet user density in these regions, and a similar rapid increase in the number of ADS users. This indicates that the new potential is being used and in this sense there is a bridging happening of the ``Digital Divide'', at least from the ADS perspective.

\begin{figure}[!ht]
  \plotone{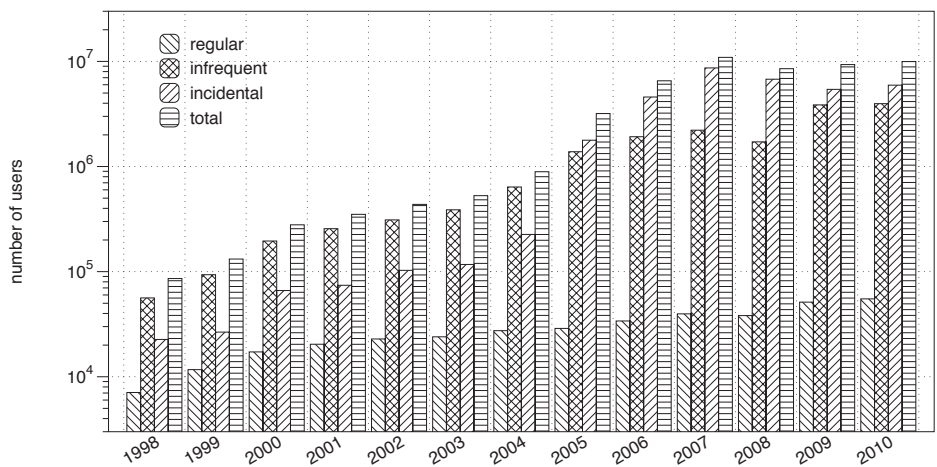}
  \caption{The number of ADS users of various types over time}
  \label{fig:ADSusers}
\end{figure}

Another metric for impact is the level of penetration in the scientific community. In other words: how many people are using the ADS regularly (10 or more times per month)? Figure~\ref{fig:ADSusers} shows that this number of regular users is still increasing.
In \citet{henneken09}~we showed that usage by regular ADS users has a median that is fairly constant at about 21 reads per month. This is an indication for the fact that all frequent ADS users on average use the ADS on a daily basis. Initially this meant that all professional astronomers use the ADS daily, but by now there must be a growing group of professionals outside astronomy (for example physicists and engineers). This is also indicated by the fact that the current number of frequent ADS users is significantly larger than the number of professional astronomers in the world (the IAU currently has just over 10,000 individual members, and there were about 17,000 different authors listed in the main astronomy journals in 2010).

Usage data also indicate another type of impact: the ADS has become a public service. ADS usage has changed qualitatively over time. The distribution of reads over users has changed, specifically the ratio of frequent to infrequent users has changed considerably over time. We feel that the strong increase in infrequent users has an impact on the science education of the general public. Between 35\% and 40\% of all ADS use actually comes from links external to the ADS. The Google, Google Scholar and Bing search engines are the largest sources, but the ADS is also linked to by thousands of static pages. For instance, Wikipedia has more than 17,000 links to the ADS. While some of the page views are scientists using Google to find a reference, the vast majority are generated by the general public. 

\subsection{Diversification and Expansion of Holdings}
In order to stay relevant for its core users, the ADS holdings must accurately reflect the complexity of the fields these people are working in. During the lifetime of the ADS, new fields emerged in astronomy and astrophysics, and existing fields became more complex, reflected in more diffuse boundaries with fields that historically had a tenuous connection with astronomy at best. The holdings of the ADS evolved accordingly. The number of journals has increased significantly over time and we currently have over 1.8 million records in our astronomy database, distributed over more than 4,500 journals. 

Our digitization efforts are another aspect of the diversification of the ADS holdings. All the major astronomy journals have been scanned back to volume 1, and they have recently been re-scanned to capture grey scale and color content. Now the ADS is focusing on scanning publications with high scientific impact and that are not otherwise available online. We have also been collaborating with librarians and observatories to digitize series of historical publications that are difficult to locate and obtain. The impact of this effort is considerable: were it not for their availability in ADS, much of this content would be simply out of the reach of researchers, librarians, and the general public. Consider, for instance, the content that ADS has digitized in collaboration with the CfA library, which consists of almost 900,000 pages of historical observatory publications. During 2009 alone, more than 1 million articles were downloaded from this collection. In other words, many historical publications have been given a new life thanks to the digitization efforts of the ADS. That this historical material is being read can be illustrated by looking at the obsolescence function of reads. The ADS logs from March and April of 2011 indicate a reads rate for historical publications of about 1.2 reads per paper per year, which agrees with the results found in \citet{kurtz05b}. 

\subsection{Trends in the Electronic Publication Process}
\begin{figure}[!ht]
  \plotone{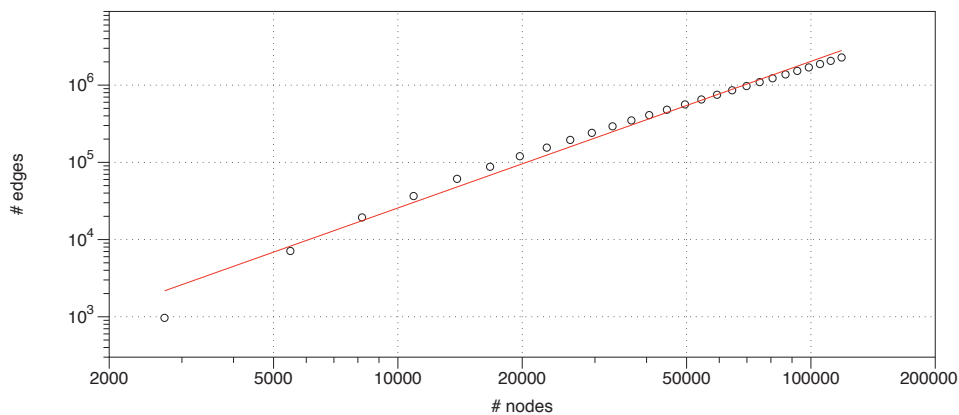}
  \caption{Number of nodes versus the number of edges in the citation network of the major astronomy journals in the period 1980 - 2006}
  \label{fig:densification}
\end{figure}

It seems almost unavoidable that the presence of the ADS would also have had an impact on the actual products of the scientific community, specifically publications. Ease of access is very likely to have an impact on the actual writing process, in the sense that publications can probably be classified as being ``published before the introduction of the World Wide Web'' and ``published after the introduction of the World Wide Web''. One way of finding classifications is to represent publications as nodes in an relational network. An obvious example is the citation network: articles A and B are directionally connected when ``A cites B''. This network turns out to have a very specific trend: high densification over time, i.e. non-linear growth. This is shown in figure~\ref{fig:densification}, showing the relation between the number of nodes and the number of edges in the network.

\begin{figure}[!ht]  
  \plotone{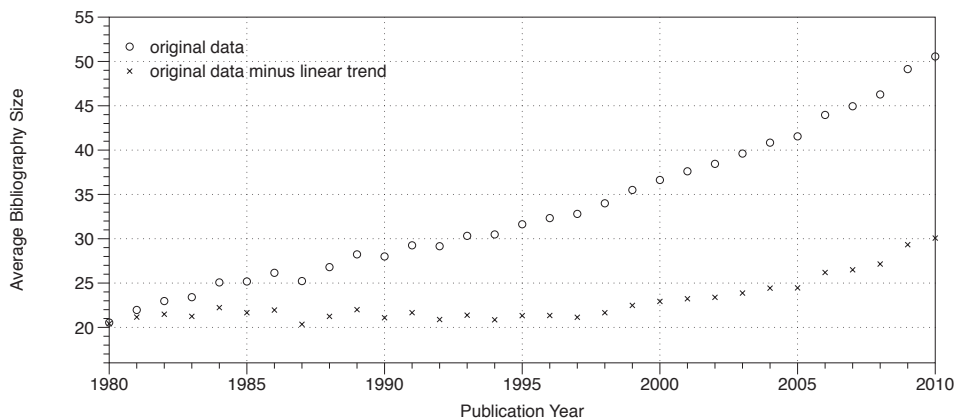}
  \caption{The average number of references in bibliographies in the main astronomy journals}  
  \label{fig:bibliographies}
\end{figure}

One implication is that, in an average sense, bibliographies have increased in length over time. This becomes abundantly clear in figure~\ref{fig:bibliographies}, which shows the average number of references in bibliographies as a function of publication year.

When an overall linear trend is subtracted (represented by ``x'' in figure~\ref{fig:bibliographies}), a deviation from this trend is observed from the mid to late 1990’s on. One could argue that the ease of access, offered by web services like the ADS, results in more citations (on average). Clearly, the ADS is not the only source for bibliographic information, but since our data set consisted solely of the major astronomy journals and the ADS has 100\% penetration in the astronomical community, it is very likely that the ADS contributed significantly to the observed trend.

The ADS also has had impact on the study of trends in the (electronic) publication process and in how literature is being used. Michael J. Kurtz has contributed significantly to our understanding of e.g. usage and citation bibliometrics (see e.g. \citet{kurtz00}~and \citet{kurtz05b}, \citet{kurtz05c}) and the discussion of the the influence of Open Access (see e.g. \citet{kurtz07}~and \citet{henneken07b}). 

\section{The Future ADS}
How is the ADS moving into the future? According to some, the Information Age is over and we are moving into the Imagination Age, where creativity and imagination are becoming the primary creators of economic value, as opposed to thinking and analysis (see e.g. King 2007). Whatever ``age'' we are in, there will remain a desire with individuals to be able to transfer information freely and to have instant access to information that would have been difficult (or even impossible) to retrieve previously. But there is definitely a new trend. It is becoming more and more common that we are faced with data collections of such magnitude and complexity, that conventional data and information discovery models brake. We need innovative ways to explore an enormous, rapidly expanding data universe. Not too long ago gigabytes seemed like a lot. The Panoramic Survey Telescope and Rapid Response System (Pan-STARRS) project is expected to produce several terabytes of raw data per night! The ADS holdings will never come close to these amounts of data, but in our case it is the complexity of the data that requires our attention, and also the more complicated demands (and perhaps expectations) of people using the ADS. Therefore, in order to stay relevant for its end-users, the ADS has to innovate and develop new ways to explore the literature universe. We are doing exactly that in our test environment ``ADS Labs'', where we expose our users to new technologies and prototype services. For example, the ``streamlined search'' allows users to find publications that are review papers for the subject they are interested in. Where in the ``classic'' approach the user had to have knowledge about what he/she is looking for and how to find it, the new streamlined search in ``ADS Labs'' offers a means to specify beforehand what a user is looking for. In addition to this, the results will be displayed in a different way. We use faceted filtering allowing our users to explore the literature by filtering collections of records by a particular property or set of properties. This is an efficient way to quickly focus on a particular subset of records, from the results of a broad search. In this way, we provide the user with a custom information environment. We feel that this approach results in an even more efficient information discovery environment than the classic approach. In addition to this, the abstract page now also includes recommendations. These recommendations are based on publication similarity, in combination with article usage information from people who use the ADS frequently. The inclusion of recommendations to the usual citations and references links adds an element of serendipity to the usual activity of searching and browsing the literature. In addition to the new abstract search, ADS Labs also offers a full text search that includes current full text for all main astronomy journals. This mode of searching add a whole new dimension to information discovery, hitherto not available. 

\section{Concluding Remarks}
We discussed a number of metrics for the impact of the ADS. These metrics are indicators for the impact of the ADS has on efficiency of information retrieval (essential for both research and scholarly communication), popularization of astronomy and the publication process itself (both in production and understanding). On all levels, the impact of the ADS has been significant. This impact has also been expressed in the form of recognition by peer organizations and prizes bestowed upon ADS staff.

It has been widely recognized (\citet{fabbiano10}) that in this era of data-intensive science, it is critical for researchers to be able to seamlessly move between the description of scientific results, the data analyzed in them, and the processes used to produce them. As observations, derived data products, publications, and object metadata are curated by different projects and archived in different locations, establishing the proper linkages between these resources and describing their relationships becomes an essential activity in their curation and preservation. The ADS, in collaboration with the VAO, the NASA archives, and the SIMBAD project, is leading the effort of better integrating and linking the research literature with the body of heterogeneous astronomical resources in the VO, allowing users as well as applications to easily cross boundaries between archives. This endeavor has been named Semantic Interlinking of Astronomical Resources (\citet{accomazzi11}), and is funded by the Virtual Astronomical Observatory Data Curation \& Preservation project. By maintaining its traditional role, but introducing innovations in its querying capabilities, and by taking on this new role in the inter-linking of information sources, the ADS intends to keep playing a central, pivotal role within the astronomical community.


\end{document}